\documentclass[onecolumn,showpacs,nobibnotes,nofootinbib]{revtex4}
\usepackage{amsmath,amssymb}
\usepackage{graphicx,subfigure}

\newcommand{\St}{{\tilde{S}}}
\newcommand{\Sta}{{\tilde{S}}^{(a)}}
\newcommand{\Tt}{{\tilde{T}}}
\newcommand{\Tta}{{\tilde{T}}^{(a)}}

\newcommand{\x}{\mathbf{x}}
\newcommand{\vk}{\mathbf{k}}
\newcommand{\vp}{\mathbf{p}}

\newcommand{\abs}[1]{\left|#1\right|}

\begin{document}

\title{Traveling waves in discretized Balitsky-Kovchegov evolution}
\author{C. Marquet}
\email{marquet@spht.saclay.cea.fr}
\author{R. Peschanski}
\email{pesch@spht.saclay.cea.fr}
\author{G. Soyez\footnote{on leave from the fundamental theoretical physics 
group of the University of Li\`ege.}}
\email{gsoyez@spht.saclay.cea.fr}
\affiliation{SPhT \footnote{URA 2306, unit\'e de recherche associ\'ee au CNRS.}, 
CEA Saclay, B\^{a}t. 774, Orme des Merisiers, 
F-91191 Gif-Sur-Yvette Cedex, France}
\author{A. Bialas}
\email{bialas@th.if.uj.edu.pl}
\affiliation{Institute of Physics, Jagellonian University, Reymonta 4, 30-059 Krakow, Poland}

\begin{abstract}
We study the asymptotic solutions of a version of the Balitsky-Kovchegov evolution with discrete steps in rapidity. We derive a closed iterative equation in momentum space. We show that it possesses traveling-wave solutions and extract their properties. We find no evidence for chaotic behaviour due to discretization.
\end{abstract}

\maketitle

\section{Introduction}\label{sec:intro}

Following the discovery of the rapid rise of the proton structure function at HERA ten years ago \cite{hera}, a great effort has been made to theoretically grasp the high-energy limit of QCD. Within the large-$N_c$ limit and in the leading-logarithmic approximation, Balitsky and Kovchegov have proposed an equation \cite{BK} combining the linear Balitsky-Fadin-Kuraev-Lipatov (BFKL) growth \cite{BFKL} and the saturation effects due to multiple scattering and gluon recombination \cite{sat}. 

This evolution to high rapidities $Y=\log(s)$ is expressed in the colour-dipole framework. Actually, one can view the evolution process as a dipole cascade, associated with a classical branching with a probability distribution given by the BFKL kernel. These branching processes appear frequently in statistical mechanics and are well suited to be considered with discrete time steps. In the case of the Balitsky-Kovchegov (BK) equation, rapidity plays the role of time and its discretization, acting as a veto in rapidity, can {\em e.g.} take into account next-to-leading-order effects \cite{veto1,veto2,veto3}. 

It has been argued recently \cite{chaos} that discretization of the BK equation may lead to chaotic behaviour of the scattering amplitude at high energy. This discretization is motivated by the fact that fluctuations around the mean-field (BK) solution can modify the scattering amplitude. This effect can be taken into account by discretizing the number of emitted gluons \cite{imm}. The evolution can then be considered as a discrete process where the emission of a gluon occurs over a rapidity interval $\Delta Y$. Our aim is to study this problem using a discretization scheme corresponding to an explicit coordinate-dependent branching process (see further eq. \eqref{eq:evolcoord}).

Recently, a discretized version of the BK equation \cite{bp} has been proposed, based on its holomorphic separability which is a crucial property of the BFKL kernel in the leading-logarithmic approximation. In addition, the BK equation has been shown \cite{mp} to admit traveling-wave asymptotic solutions. It means that, when rapidity increases, the dipole density $T(r,Y)$ depends on the dipole size $r$ and the rapidity $Y$ only through the variable $rQ_s(Y)$ where the so-called {\em saturation scale} $Q_s^2(Y)\propto \exp(\bar\alpha v_c Y)$ is related to the speed $v_c$ of the wave. Both the speed and the tail of the front $T(rQ_s(Y))$ can be deduced from the BFKL kernel.

%In addition, the BK equation has been shown \cite{mp} to admits traveling-wave asymptotic solutions. It means that, when rapidity increases, the dipole density $T(r,Y)$ depends on the dipole size $r$ and the rapidity $Y$ only through the variable $rQ_s(Y)$ where both the rapidity dependence of the so-called saturation scale $Q_s(Y)\approx v_c Y$ and the form of the tail of the front can be deduced from the BFKL kernel.

In this paper, we show analytically and numerically that the discretized version of the BK equation leads to the formation of traveling waves and we study their properties. We shall show that the velocity of the wave, {\em i.e.} the exponent of the saturation scale, increases w.r.t. the continuous equation and the slope of the front decreases.

Finally, in another study of a discretized version of the BK equation \cite{chaos}, the authors have suggested the apparition of chaos in the evolution. We shall discuss that point using our discretization scheme.

The paper is organised as follows: in section \ref{sec:bk} we present the discrete-time formulation of the BK evolution equation, we continue with its solutions and, in particular, the traveling-wave solutions in section \ref{sec:sol}. We then finish by a conclusion and a discussion of our results.

\section{Discretized evolution equation}\label{sec:bk}

The impact-parameter-independent Balitsky-Kovchegov (BK) equation for the $S$-matrix element corresponding to the dipole-target scattering amplitude can be written
\begin{equation}\label{eq:bk}
\partial_Y S(\x_{01},Y) = \int d^2 x_2\, K(\x_0, \x_1; \x_2)
                          \left[ S(\x_{02},Y)S(\x_{21},Y)-S(\x_{01},Y)\right],
\end{equation}
where $\x_{ij} = \x_i-\x_j$ are the transverse coordinates of the dipoles, $x_{ij}=|\x_{ij}|$, $r=x_{01}$, and 
\[
K(\x_0, \x_1; \x_2)\,d^2 x_2 = \frac{\alpha_s N_c}{2\pi^2} \frac{x_{01}^2}{x_{02}^2 x_{21}^2}\,d^2 x_2 
\]
is the kernel giving the probability for a dipole of coordinates $(\x_0,\x_1)$ to split into two child dipoles of coordinates $(\x_0,\x_2)$ and $(\x_2,\x_1)$. The BFKL kernel $K$ has the property of {\em holomorphic separability}, which means that if we introduce
\[
z = \frac{(\x_{02})_1+i(\x_{02})_2}{x_{01}^2},\quad \bar z = \frac{(\x_{02})_1-i(\x_{02})_2}{x_{01}^2},
\]
the kernel factorises into a function of $z$ times a function of $\bar{z}$.

The discretized version of the evolution kernel is fixed to keep this property\footnote{To obtain the regularised kernel $K_d$, one replaces $\scriptstyle{\frac{x_{01}^4}{x_{02}^2 x_{21}^2}}$ by $\scriptstyle{\left(\frac{x_{01}^4}{x_{02}^2 x_{21}^2}\right)^{1-a}}$ and normalises. This amounts to replace $K$ by $NK_d$ in \eqref{eq:bk}.}:
\begin{equation}\label{eq:Kd}
K_d(\x_0, \x_1; \x_2) \,d^2 x_2 = \frac{1}{N}\frac{\alpha_s N_c}{2\pi^2} 
   \left(\frac{x_{01}^4}{x_{02}^2 x_{21}^2}\right)^{1-a}\,
   \frac{d^2 x_2}{x_{01}^2} 
\end{equation}
with
\begin{equation}\label{eq:norm}
N = \int \frac{d^2 x_2}{x_{01}^2} \,\frac{\alpha_s N_c}{2\pi^2} 
   \left(\frac{x_{01}^4}{x_{02}^2 x_{21}^2}\right)^{1-a}\,
  = \frac{\alpha_s N_c}{2\pi}\, \frac{\Gamma^2(a)\,\Gamma(1-2a)}{\Gamma^2(1-a)\,\Gamma(2a)}
\end{equation}
where the regulator $a$, chosen between 0 and $1/2$, allows to separate the real and virtual contributions in \eqref{eq:bk} which gives for a rapidity-discretized amplitude
\[
\partial_Y S(\x_{01})\longrightarrow\frac{S_{n+1}(\x_{01})-S_n(\x_{01})}{\Delta Y} = N \int d^2 x_2 \, K_d(\x_0, \x_1; \x_2) \, S_n(\x_{02}) S_n(\x_{21}) - NS_n(\x_{01}).
\]
One thus fixes the rapidity step $\Delta Y=1/N$ such that
\[
Y = n\Delta Y = \frac{n}{N}\qquad \text{ with }\quad n \in {\rm I\! N},
\]
and the evolution equation for the discretized case takes the iterative form
\begin{equation}\label{eq:evolcoord}
S_{n+1}(\x_{01}) = \int d^2 x_2 \, K_d(\x_0, \x_1; \x_2) \, S_n(\x_{02}) S_n(\x_{21}).
\end{equation}
This evolution corresponds to a branching process with a probability distribution given by the kernel $K_d$.

Let us derive the evolution equation in momentum space. For this sake, we introduce
\begin{equation}\label{eq:sta}
\Sta(\vk) \equiv 1-\Tta(\vk) = \frac{\int d^2x\,e^{i\vk.\x}\,x^{2(a-1)}\,S(\x)}{\int\,d^2x\, e^{i\vk.\x}\,x^{2(a-1)}}  = 1-\frac{\int d^2x\,e^{i\vk.\x}\,x^{2(a-1)}\,T(\x)}{\int\,d^2x\, e^{i\vk.\x}\,x^{2(a-1)}} 
%= \frac{\Gamma(1-a)}{\Gamma(a)} 2^{1-2a} k^{2a} \int dx\, \BesselJ{0}{kx} x^{2a-1}S(x),
\end{equation}
where both $\Tta(\vk)$ and $\Sta(\vk)$ are between 0 and 1. Equation \eqref{eq:evolcoord} becomes \cite{bp}
\begin{equation}\label{eq:evolimp}
\St^{(2a)}_{n+1}(\vk) = \left[ \Sta_n(\vk) \right]^2 \qquad\text{ or }\quad 
\Tt^{(2a)}_{n+1}(\vk) = 2 \Tta_n(\vk) - \left[ \Tta_n(\vk) \right]^2.
\end{equation}

For the purpose of our studies, a closed equation involving only $\Tta$ is required.
To achieve this, one first invert \eqref{eq:sta} leading to
\begin{equation}\label{eq:statos}
T(\x) = \frac{\Gamma(a)}{\Gamma(1-a)} \frac{x^2}{4\pi} \int d^2k\,e^{-i\vk.\x} 
        \left(\frac{4}{k^2x^2}\right)^a \Tta(\vk)
      = \frac{\Gamma(2a)}{\Gamma(1-2a)} \frac{x^2}{4\pi} \int d^2k\,e^{-i\vk.\x} 
        \left(\frac{4}{k^2x^2}\right)^{2a} \Tt^{(2a)}(\vk),
\end{equation}
where the second equality is simply obtained by replacing $a$ by $2a$. Inserting this into equation \eqref{eq:sta} allows to express $\Tta_{n+1}(\vk)$ in terms of $ \Tt^{(2a)}_{n+1}(\vk)$. The evolution equation \eqref{eq:evolimp} finally takes the form
\begin{equation}\label{eq:evolclosed}
\Tta_{n+1}(\vk) 
  = \frac{\Gamma(1-a)\,\Gamma\left(a+\frac{1}{2}\right)}
         {\Gamma(a)\,\Gamma\left(\frac{1}{2}-a\right)}
    \frac{2^{4a}}{2\pi}
    \int \frac{d^2p}{(\vk-\vp)^2}\left[\frac{k^2(\vk-\vp)^2}{p^4}\right]^a
         \left\{2 \Tta_n(\vp) - \left[ \Tta_n(\vp) \right]^2\right\}.
\end{equation}

It is clear from \eqref{eq:evolclosed} that if $\Tta_n(\vp)$ depends only on the modulus $p$ of $\vp$, $\Tta_{n+1}(\vp)$ also features that property. Thus, starting with an angular-independent initial condition $\Tta_0(p)$, this property is conserved throughout the evolution. Hence, one can integrate out the angular $\varphi$ dependence in \eqref{eq:evolclosed}. Introducing the integration variable $\sin^2(\varphi/2)$, one obtains
%If one considers that $\Tta(\vp)$ depends only on the modulus $p$ of $\vp$ one can integrate out the angular $\varphi$ dependence. Introducing the integration variable $\sin^2(\varphi/2)$, one obtains
\begin{equation}\label{eq:evol}
\Tta_{n+1}(k) 
 = 2^{4a} \frac{\Gamma(1-a)\,\Gamma\left(a+\frac{1}{2}\right)}
               {\Gamma(a)\,\Gamma\left(\frac{1}{2}-a\right)}
   \int_0^\infty \frac{p\,dp}{(k+p)^2}
        \left[\frac{k^2(k+p)^2}{p^4}\right]^a
        \!\!\phantom{F}_2F_1\left(\frac{1}{2},1-a;1;\frac{4kp}{(k+p)^2}\right)
	\left\{2 \Tta_n(p) - \left[ \Tta_n(p) \right]^2\right\}.
\end{equation}

Before considering the solutions of this process, let us show how one can reconstruct the physical amplitude from $\Tta$. The physical $T$-matrix in momentum space is defined as usual by
\begin{equation}
\Tt(k) = \frac{1}{2\pi} \int \frac{d^2x}{x^2}\, e^{i\vk.\x}\, T(x).
\end{equation}
If we combine this with equation \eqref{eq:statos}, one gets
\begin{equation}\label{eq:ttatot}
\Tt_n(k) 
 = \int_0^\infty \frac{p\,dp}{(k+p)^2}
        \left[\frac{(k+p)^2}{p^2}\right]^a
        \!\!\phantom{F}_2F_1\left(\frac{1}{2},1-a;1;\frac{4kp}{(k+p)^2}\right)
	\Tta_n(p).
\end{equation}
%Note that the integration kernel involved in the reconstruction of the physical amplitude can be recast in the same form as the kernel used for the branching process by using $k^{2a}\Tta(k)$ and $k^{2a}\Tt(k)$.

Equation \eqref{eq:evol}, together with the transformation \eqref{eq:ttatot}, constitute a closed iterative description of the discretized evolution.

\section{Asymptotic properties: traveling-wave solutions}\label{sec:sol}

\begin{figure}
\subfigure[Rapidity step]{\includegraphics[scale=0.66]{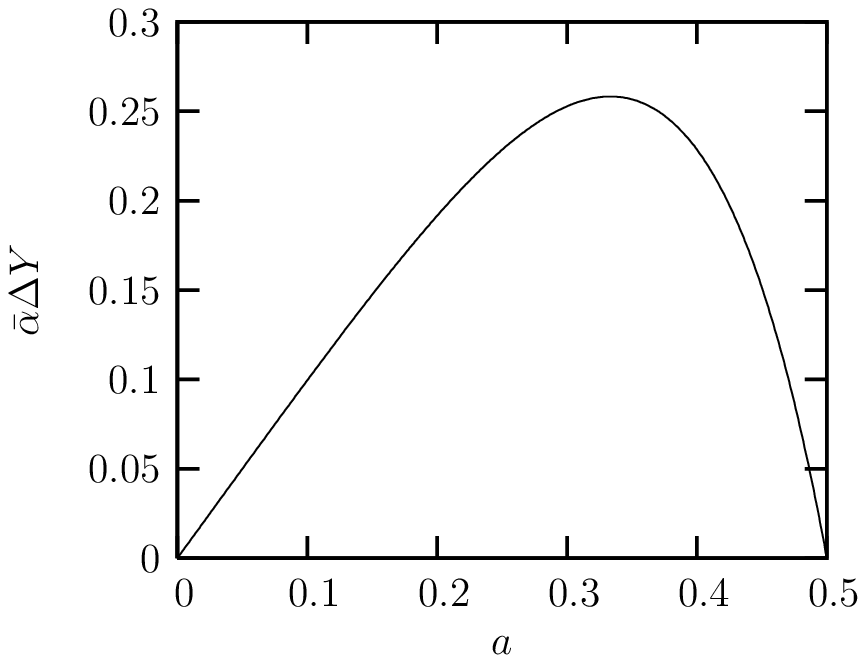}}
\subfigure[Critical slope]{\includegraphics[scale=0.66]{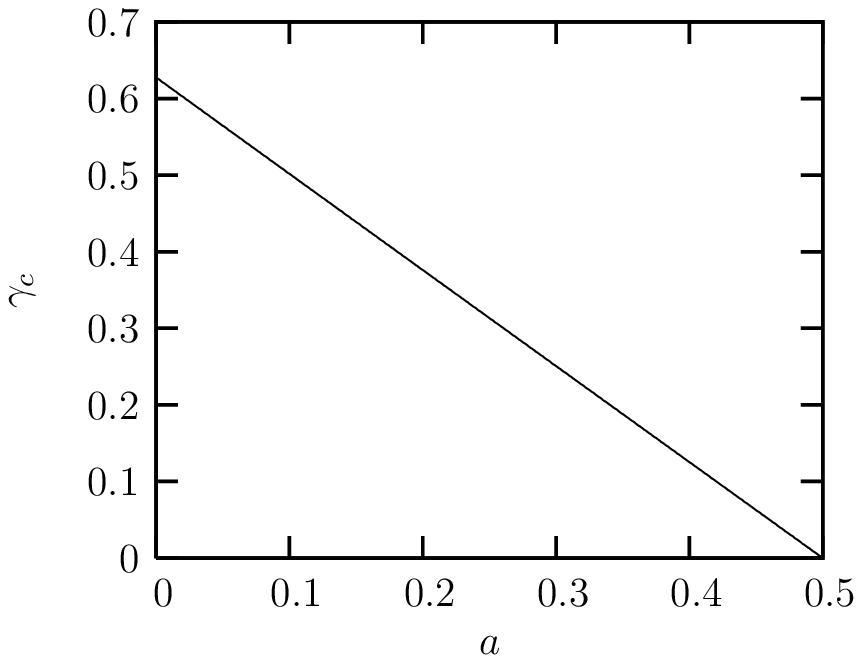}}
\subfigure[Critical velocity (crosses: numerical results)]{\includegraphics[scale=0.66]{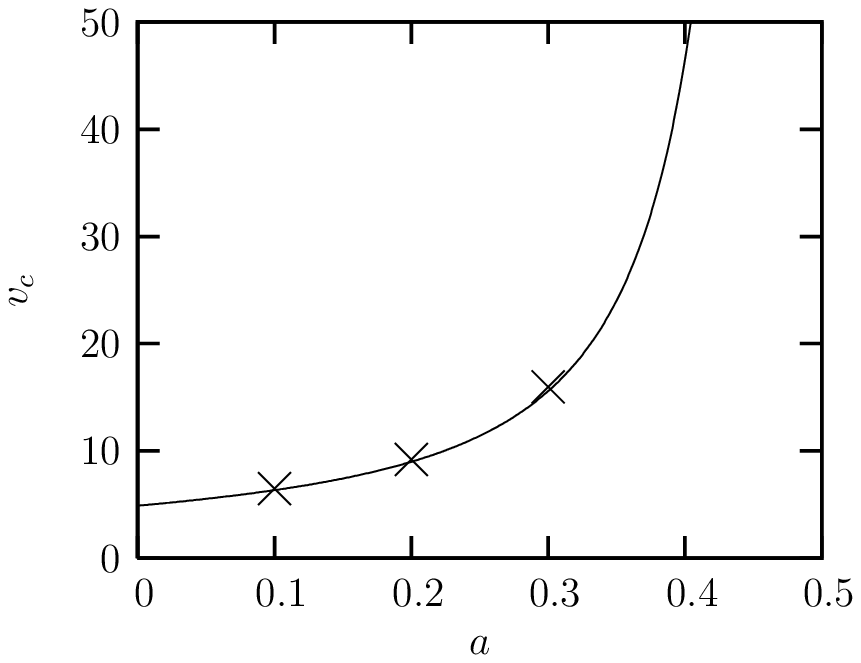} \label{fig:speedth}}
\caption{Theoretical predictions for the traveling-wave properties as a function of the regulator $a$.}\label{fig:pred}
\end{figure}

In this section, we discuss the solutions of the branching process \eqref{eq:evol} and compare them with the properties of the Balitsky-Kovchegov equation \eqref{eq:bk}. 

To begin with, let us recall the important properties of the asymptotic solutions of the BK equation. It has been observed recently \cite{mp} that the BK equation admits {\em traveling waves} as asymptotic solutions: at large rapidities, the amplitude takes the form of a wave of speed $v_c$
\[
\Tt(k,Y) = \Tt\left[\log(k^2)-\bar\alpha v_c Y \right].
\]
In physical terms, this means that $\Tt$ depends only on the ratio $k/Q_s(Y)$, {\em i.e.} satisfies the {\em geometric-scaling} property, where the {\em saturation scale} is given by $Q_s^2\propto\exp(\bar\alpha v_c Y)$. 

One can even go beyond this result, valid for infinitely large $Y$, and compute pre-asymptotic corrections to the speed of the wave. Expressed in terms of the saturation scale, one obtains \cite{mp}
\begin{equation}\label{eq:qs}
\partial_Y \,\log\left[Q_s^2(Y)\right] 
  = \bar\alpha v_c - \frac{3}{2\gamma_c}\frac{1}{Y} 
  + \frac{3}{2\gamma_c^2}\sqrt{\frac{2\pi}{\bar\alpha\chi''(\gamma_c)Y^3}}
  + {\cal {O}}\left(\frac{1}{Y^2}\right),
\end{equation}
where $\chi(\gamma)$ is the eigenvalue of the BFKL kernel and the critical parameters $\gamma_c$ and $v_c$ are determined by
\begin{equation}\label{eq:crit}
v_c = \chi'(\gamma_c) = \frac{\chi(\gamma_c)}{\gamma_c}.
\end{equation}
In addition, at large rapidities, the tail of the front is given by
\begin{equation}\label{eq:tail}
\Tt(k) \underset{Y\to\infty}{=}
\log\left(\frac{k^2}{Q_s^2(Y)}\right)\,
\abs{\frac{k^2}{Q_s^2(Y)}}^{-\gamma_c}\,
\exp\left[-\frac{1}{2\bar\alpha\chi''(\gamma_c)Y}\log^2\left(
\frac{k^2}{Q_s^2(Y)}\right)\right].
\end{equation}

Now that we have summarised the properties of the asymptotic solutions of the BK equation, let us consider the case of the discretized equation \eqref{eq:evolcoord}. More precisely, we shall prove that one can also infer traveling-wave solutions in this equation. 

In order to achieve this task, we have to show \cite{mp} that the solution of the linear equation  ($T=1-S$)
\[
T_{n+1}(\x_{01}) = \int d^2x_2\,K_d(\x_0,\x_1;\x_2)\left[T_n(\x_{02})+T_n(\x_{21})\right]
\]
can be expressed as a superposition of waves. Going to the Mellin representation
\[
T_n(x^2) = \int \frac{d\gamma}{2i\pi}\,(x^2)^\gamma\,{\cal{T}}_n(\gamma)
\]
one easily gets
\[
{\cal{T}}_n(\gamma) = {\cal{T}}_0(\gamma)\,\exp\left[N\chi^{(a)}(\gamma)Y \right]
\]
with
\begin{equation}
\chi^{(a)}(\gamma) = \log\left[
2\,\frac{\Gamma(a+\gamma)\,\Gamma(1-a)\,\Gamma(1-2a-\gamma)\,\Gamma(2a)}
{\Gamma(a)\,\Gamma(1-a-\gamma)\,\Gamma(1-2a)\,\Gamma(2a+\gamma)}\right].
\end{equation}
The kernel $\chi^{(a)}(\gamma)$, which is the logarithm of the Mellin transform of the kernel $K_d$ (see \eqref{eq:Kd}), describes the linear growth, as does the BFKL kernel $\chi(\gamma)$ in the continuous case. One can thus predict that equation \eqref{eq:evol} admits traveling waves as asymptotic solution with the critical exponent $\gamma_c$ and the critical speed $v_c$ fixed by equation \eqref{eq:crit} with $N\chi^{(a)}(\gamma)$ instead of $\chi(\gamma)$.
The rapidity dependence of the saturation scale and the tail of the front are given by equations \eqref{eq:qs} and \eqref{eq:tail}, where the critical parameters are now obtained from the kernel $N\chi^{(a)}(\gamma)$:
\begin{equation}\label{eq:crita}
v_c = N[\chi^{(a)}]'(\gamma_c) = \frac{N\chi^{(a)}(\gamma_c)}{\gamma_c}.
\end{equation} 

In figure \ref{fig:pred}, we represent the main properties of the traveling waves as a function of the regulator $a$. The value of the rapidity step (or, more precisely, $\bar\alpha\Delta Y$) shows a maximum of 0.258 for $a\approx 1/3$. When $a$ goes from 0 to 0.5, the asymptotical slope of the wave, given by $\gamma_c$, decreases\footnote{Up to a small correction of order $10^{-4}$, we observe that $\gamma_c(a) \simeq \gamma_c(0)(1-2a)$.} while its speed increases. In the limit $a\to 0$, one recovers the parameters from the BK equation: $\gamma_c \approx 0.6275$ and $v_c\approx 4.88$.

%\begin{figure}
%\begin{center}
%\includegraphics[scale=0.7]{front-1.8.eps}
%\end{center}
%\caption{Formation of a traveling wave as rapidity increases. The different curves correspond, from left to right, to $Y\approx 12n$, $n=0,\dots,10$ and we have taken $a=0.1$.
%%$Y=0, 11.94, 23.88, 35.82, 47.76, 59.70, 71.65, 83.59, 95.53, 107.5, 119.9$.
%}\label{fig:front}
%\end{figure}

\begin{figure}
\begin{center}
\includegraphics[scale=0.7]{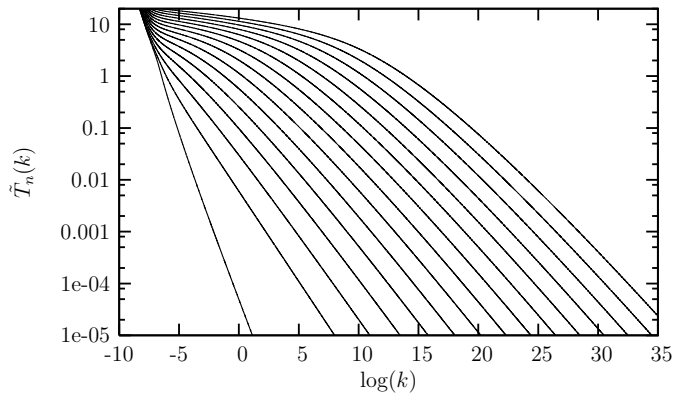}
\end{center}
\caption{Traveling-wave pattern from discretized evolution. We have fixed $a=0.3$ and $\bar\alpha=0.2$, corresponding to rapidity steps of $\Delta Y=1.264$. The amplitude $\Tt_n(k)$ is shown, from left to right, for $n=0,1,\dots,14$.}\label{fig:steps}
\end{figure}

\begin{figure}
\begin{center}
\includegraphics[scale=0.7]{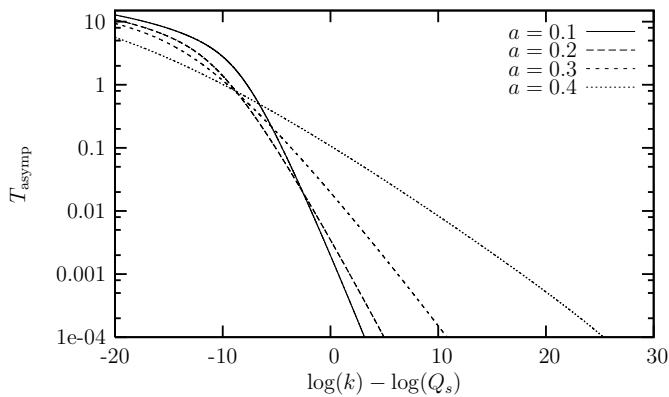}
\end{center}
\caption{Comparison of the traveling-wave front for different values of the regulator $a$. We observe that the slope decreases as $a$ increases.}\label{fig:slope}
\end{figure}

\begin{figure}
\begin{center}
\includegraphics[scale=0.7]{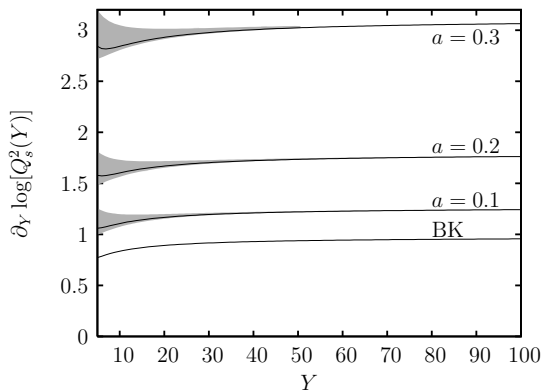}
\end{center}
\caption{Speed of the traveling wave. The solid curve shows the theoretical prediction \eqref{eq:qs} for $\partial_Y \log(Q_s^2)$. The gray area corresponds to numerical computation using different extractions of the saturation scale (see text, eq. \eqref{eq:extractqs}): the lower(higher) border corresponds to $\Tt(Q_s)=0.1$ ($0.001$). The theoretical result for the continuous case has been added for comparison.}\label{fig:speed}
\end{figure}

In order to check those predictions, we have performed a numerical analysis of equation \eqref{eq:evol}. We start with a initial condition for the physical amplitude given in coordinate space by
\[
S(x) = \exp\left(-x^\rho\right),
\]
where the parameter $\rho$ is chosen such that both $S(x)$, $\Tta(k)$ and $\Tt(k)$ are positive. Changing $\rho$ allows us to discuss the effect of the initial condition on the formation and properties of the traveling wave. From this initial condition, one can compute $\Tta_0(k)$ using $\eqref{eq:sta}$. The evolution is then performed by iterating the branching process \eqref{eq:evol} and the physical amplitude is recovered through \eqref{eq:ttatot}. The value of the strong coupling constant has been kept fixed $\bar\alpha = 0.2$.

First of all, let us discuss the formation of the traveling wave. To emphasise clearly the effect of discretization, we have chosen $a=0.3$, $\bar\alpha=0.2$ and\footnote{The choice of the initial condition is due to physical constraints, imposing a finite range. For example, for $\rho=2$ the Fourier-transformed amplitude $\Tta(k)$ is not positive definite, while $\rho=1$ corresponds to a subcritical slope which is not the case in QCD.} $\rho=1.2$, leading to a large rapidity step $\Delta Y = 1.264$ (for $a=0.1$, this gives $\Delta Y\approx 0.5$, see figure \ref{fig:pred}). The result of the evolution is plotted in figure \ref{fig:steps} where each curve corresponds to one step in rapidity {\em i.e.} to $\Tt_0(k),\dots,\Tt_{14}(k)$. We obviously see from the figure that the trace of the initial condition is lost during the evolution and that a traveling wave is formed. One can also show that if the initial condition is not steep enough, the slope of the initial condition is preserved by the evolution as for the continuous case.

If we consider different values of the regulator $a$, the slope of the resulting traveling wave, given by $\gamma_c$ provided the initial condition is steep enough, shall depend on $a$. This slope, extracted from \eqref{eq:crita}, is expected to decrease when $a$ increases. In figure \ref{fig:slope}, we show the form of the front obtained after evolution for different values of $a$, and it clearly appears that the slope decreases significantly for large values of $a$.

Finally, we have to check that the rapidity dependence of the saturation scale is in agreement with equation \eqref{eq:qs}. To extract the saturation scale from the numerical simulations we solve
\begin{equation}\label{eq:extractqs}
\Tt(Q_s(Y),Y) = \Tt_0.
\end{equation}
The level $\Tt_0$ at which we extract the saturation scale is not fixed, but it only introduces subleading corrections to the saturation scale when rapidity increases. In our analysis, we have chosen $\Tt_0$ between 0.001 and 0.1, which is characteristic for the region where the traveling wave is formed. We can then plot the numerical value for the saturation scale together with the predictions from \eqref{eq:qs}. The result is shown in figures \ref{fig:speedth} and \ref{fig:speed} for different values of the regulator $a$ and for $\rho=1.2$, together with the theoretical prediction for the continuous BK equation. In addition to the perfect agreement between the theoretical expectation and the numerical results, we observe that the speed of the traveling wave becomes larger and larger as $a$ increases.

\section{Discussion and conclusion}\label{sec:ccl}

Throughout this paper, we have seen that the property of holomorphic separability gives a natural way to discretize the BK equation in rapidity through the introduction of a regulator $a$ which allows to adjust the rapidity step. The evolution then turns into a discrete branching process which can be cast in a closed form \eqref{eq:evol} from which we can easily recover the physical dipole amplitude \eqref{eq:ttatot}. 

We have studied the asymptotic properties of this discretized equation both analytically and numerically. An analysis of the new evolution in Mellin space predicts the formation of traveling-wave solutions as it is the case for the continuous BK equation. We have tested numerically that these traveling waves indeed appear with the correct rapidity dependence of the saturation scale $Q_s^2$ and the correct behaviour of the front. The most important effects of discretization is to increase the speed of the wave and to tame the $k^2$-dependence of the tail of the front when $a$ increases ($a\to 0$ corresponds to the continuous limit).

It is important to mention that, in comparison with the approach proposed recently in Ref. \cite{chaos}, our evolution equation does not generate chaos. We do believe that this is due to a more careful treatment of the rapidity steps. Indeed, if one discretizes the $Y$-derivative of the BK equation using steps in $\bar\alpha Y$ of size $\delta$ and keep only the lowest order expression for the BFKL kernel $\chi(\gamma)\approx\chi(1/2)$, one gets
\[
T_{n+1} = \left[1+\delta\chi(1/2) \right] T_n - T_n^2.
\]
Defining $U = [1+\delta\chi(1/2)]^{-1} T$, the evolution takes the form of the famous logistic map \cite{logistic}
\[
U_{n+1} = \lambda U_n (1-U_n),
\]
with the Malthusian parameter $\lambda = 1+\delta\chi(1/2)$. In \cite{chaos}, the authors adopt $\delta=1$, corresponding to large rapidity steps $\Delta Y = 1/{\bar\alpha} \approx 5$, giving $\lambda=3.77$. This value being larger than the accumulation point $\lambda_{\text{ap}}\approx 3.56994567$, the evolution generates chaos. In the approach developed in this paper, the value of $\delta$ is given by $1/N$ which depends on $a$. One has noticed that the maximal value of the step size is $\delta\approx 0.258$, reached for $a\approx 0.333$. This corresponds to a Malthusian parameter of $1.715$. Since this is smaller than $3$, we lay in the $1$-cycle region and no chaos is observed. Note finally that our evolution equation \eqref{eq:evolimp}, which does not require any approximate expansion of the kernel, is itself very similar to the logistic map up to a specific convolution (see formula \eqref{eq:evolclosed}).

\begin{acknowledgments}
G.S. is funded by the National Funds for Scientific Research (Belgium).
\end{acknowledgments}

\end{document}